\documentclass[12pt]{iopart}
\usepackage{iopams}
\expandafter\let\csname equation*\endcsname\relax
\expandafter\let\csname endequation*\endcsname\relax
\usepackage{amsmath}

\usepackage{cite}
\usepackage{bm}
\usepackage{float}
\usepackage{mathtools}
\usepackage{multirow}

\usepackage{bbold}
\usepackage{bbm}
\usepackage{braket}

\usepackage{amsthm}
\usepackage{mathrsfs}
\usepackage[usenames]{xcolor}
\usepackage{gensymb}
\usepackage{appendix}
\usepackage{graphicx}
\usepackage[caption=false]{subfig}
\usepackage[normalem]{ulem}
\usepackage{color}

\usepackage{soul}
\usepackage{hyperref}

\newcommand{\dket}[1]{| #1 \rangle \rangle}

\def\cL{\check{\mathcal L}}
\def\mean#1{\left< #1 \right>}

\begin{document}


\title{A dissipative time crystal with or without $\mathbb Z_2$ symmetry breaking}

\author{Crist\'{o}bal Lled\'{o} and Marzena H. Szyma\'{n}ska}
\address{Department of Physics and Astronomy, University College London,
Gower Street, London, WC1E 6BT, United Kingdom}

\date{\today}

\begin{abstract}
We study an emergent semiclassical time crystal composed of two interacting driven-dissipative bosonic modes. The system has a discrete $\mathbb Z_2$ spatial symmetry which, depending on the strength of the drive, can be broken in the time-crystalline phase or it cannot. An exact semiclassical mean-field analysis, numerical simulations in the quantum regime, and the spectral analysis of the Liouvillian are combined to show the emergence of the time crystal and to prove the robustness of the oscillation period against quantum fluctuations.  
\end{abstract}



\maketitle
\section{Introduction}
The advances in preparing and manipulating quantum matter in the laboratory during the past decades has led to a growing interest in out-of-equilibrium quantum phases \cite{kinoshita2006quantum,schneider2012fermionic,rechtsman2013photonic,jotzu2014experimental, schreiber2015observation,eisert2015quantum,smith2016many,Kucsko2018PRL,mciver2020light,Editorial}. Simultaneously, a great degree of attention has been devoted to the search for the spontaneous breaking of time-translational invariance \cite{WilczekTC, bruno2013comment, wilczek2013reply, bruno2013impossibility, Sacha-DTC-Proposal, Else-DTC-Proposal, Khemani-DTC-Proposal, SachaRPP}. Both efforts has converged on the realization of a time-crystalline phase of matter, the so called discrete or Floquet time crystals \cite{Choi-DTC-Observation, Zhang-DTC-Observation}. There, a many-body quantum system self-organizes and responds with a period different from the one imposed by the time-periodic external drive, breaking the discrete time-translational symmetry. An important characteristic of Floquet time crystals is that strong disorder is needed to induce many-body localization preventing the system from absorbing energy from the drive and heating up towards a featureless thermalized state.  

Another type of time crystal dissipates energy to the environment instead of relying on disorder \cite{TC-decoherence, Keeling2018, Poletti2018, Owen2018, Pal2018, Rey2018, Ueda2018, Lesanovsky2019, Nalitov-dimer-TC-2019, Kessler2019-PRA, Zhu_2019, Lledo2019_BHD, Savona2019, Esslinger-Science2019, chiacchio2019dissipation, Buca-dissipation-2019PRL, chinzei2020TC, kessler2020continuous}. Interestingly, a subgroup of these are not driven at all \cite{Keeling2018} or the driving is such that the time-dependence can be completely eliminated by moving to a rotating frame \cite{Owen2018, Rey2018, Nalitov-dimer-TC-2019, Kessler2019-PRA, Lledo2019_BHD, Savona2019, Esslinger-Science2019, chiacchio2019dissipation, Buca-dissipation-2019PRL, kessler2020continuous}. In theses cases, the symmetry breaking is assessed with respect to the time-independent dynamical generator, usually the Lindblad superoperator $\mathcal L$ of a Markovian master equation, $\partial_t \hat \rho = \mathcal L(\hat \rho)$. Such dissipative quantum systems have one or more attractive steady states that respect the time-translational invariance of $\mathcal L$. In the thermodynamic limit, however, the steady state might never be reached, signalling that the continuous time-translational symmetry is broken. If one finds a time-periodic response on a function $f(\tau) = \lim_{t,N\to\infty} \Tr[\hat O \hat \rho(t+\tau)]$ ---$N$ being the system size--- for a suitable system operator $\hat O$ one can say that a time crystal has been formed. Crucially, the period is not constrained to integer multiples of the drive period and can vary continuously with the system's parameters.

Most of the dissipative time crystals with continuous time-translation symmetry studied so far rely on long-range interactions \cite{Keeling2018, Owen2018, Rey2018, Kessler2019-PRA, Esslinger-Science2019, chiacchio2019dissipation, Buca-dissipation-2019PRL, kessler2020continuous}, which occur naturally in systems with dipolar interactions or can be engineered, for instance, by coupling matter to a common resonant mode of a lossy cavity. This means those systems can be well described by mean-field equations in the infinite-volume thermodynamic limit, and so be regarded as emergent semiclassical time crystals. A few exceptions \cite{Lledo2019_BHD, Savona2019} rely on a different notion of 'thermodynamic limit' in effectively zero dimensions \cite{Carmichael2015}, where the number of bosonic excitations diverge in a system with one or a few bosonic modes and quantum fluctuations become negligible. These therefore belong to the same kind of time crystals.

The majority of the time crystals which have been studied so far have an underlying symmetry in addition to the time-translational symmetry, and the two are broken together. In Floquet time-crytals, for example, it is normally a global parity symmetry ($\mathbb Z_2$), and this leads to long-range correlations in both time and space. For this reason the time-crystalline order is sometimes dubbed \textit{spatio-temporal order} \cite{Else-DTC-Proposal, Khemani-DTC-Proposal, Keyserlingk_PRB_2016, Khemani_PRB_2017, russomanno2017floquet}. This spatial symmetry is not a requisite in continuous time crystals \cite{Keeling2018, Savona2019}, yet when it is present it is broken \cite{Owen2018, Rey2018, Lledo2019_BHD, Kessler2019-PRA, Esslinger-Science2019,chiacchio2019dissipation, Buca-dissipation-2019PRL, kessler2020continuous}.

In this work we study an emergent semiclassical time crystal in a dissipative system with two interacting bosonic modes which are driven and posses a minimal spatial $\mathbb Z_2$ symmetry. We show that, depending on the strength of the drive, the time-crystalline phase is either accompanied by the $\mathbb Z_2$-symmetry breaking or it is not. This only occurs in a well-defined 'thermodynamic limit' in which the numbers of bosonic excitations diverge. By analysing how our quantum system scales towards this limit, we show the emergence of these time-ordered phases (i) proving that the period of the oscillations is robust against quantum fluctuations as well as (ii) providing insight on the feasibility of observing long-lived oscillations in an experimental realization in which the system might be far from the 'thermodynamic limit'.

The system we consider is the Bose-Hubbard dimer (BHD), which in its closed system version has been a prototypical model capable of explaining interesting macroscopic coherent dynamics, such as self-trapping and nonlinear Josephson-like oscillations in a Bose-Einstein condensate trapped in a double-well potential \cite{BEC-self-trapping, Observation-BEC-self-trapping, Observation-ac-tunneling,self-trapping-Paris}. While there is an intrinsic quantum aspect to Bose-Einstein condensates --- their statistics --- quantum correlations are normally irrelevant and the condensates can be modelled by classical nonlinear wave equations. Nevertheless, by taking into account quantum correlations, it has been suggested that the Josephson-like oscillations in the \textit{driven-dissipative} BHD dimer \cite{Parametric-instability, parametric-instability-Paris} can be regarded as a signature of a dissipative time crystal \cite{Lledo2019_BHD, Savona2019}. 

Unlike the most common case in which each mode has its own dissipative channel, the BHD we consider has nonlocal dissipation (also called dissipative coupling or collective dissipation). The idea of having self-sustained periodic oscillations in Bose-Einstein condensates with this kind of dissipation has been explored in the context of weak-lasing and frequency comb generation \cite{RuboPRB2012, Nalitov-dimer-TC-2019, Hui-Deng-PRB2020, Ruiz-Chaos-BHD-PRB2020}, although with incoherent instead of coherent drive. Nonlocal dissipation is also encountered in other models of dissipative time crystals \cite{Keeling2018, Rey2018, Wang18}. We propose the intuitive explanation that such collective process enhances synchronization between different parts of the system, which is in a sense what happens in continuous time crystals \cite{Rey2018}. Even though it is often neglected in theoretical modelling, nonlocal dissipation occurs naturally in many systems in which there is one environment weakly-interacting with the whole system, being a crucial requirement for obeying quantum detailed balance \cite{BreuerBook}. In a previous work \cite{Lledo2019_BHD}, we have shown that this dissipation together with the interactions decouple the two collective modes in the coherently driven BHD, forming periodic oscillations between the two modes and thus a time crystal. There, the two bosonic modes were symmetrically driven which rendered the time crystal bistable: for a pump-strength window, it was possible to find two distinct time-crystalline periods depending on the initial condition imposed on the system. The dimer had a spatial $\mathbb Z_2$ symmetry (the swap mode 1 $\leftrightarrow$ mode 2), which was found to be broken throughout the time crystalline phase. Here, considering a different driving configuration, we show that the spatial $\mathbb Z_2$ symmetry does not need to be broken in all regions of the time crystalline phase.

\section{The model}

We consider an open BHD with nonlocal dissipation which evolves according to the Lindblad equation ($\hbar =1$)
\begin{equation} \label{equ: Lindblad equation}
\partial_t \hat \rho = -i[\hat{\mathcal H}, \hat \rho] + \gamma  \mathcal D[\hat a_1 + \hat a_2](\hat \rho),
\end{equation}
where $\mathcal D[\hat X](\hat \rho) = \hat X \hat \rho \hat X^\dag -(1/2)(\hat \rho \hat X^\dag \hat X + \hat X^\dag \hat X \hat \rho)$ is the standard Lindblad dissipator. In a frame rotating with the pump frequency $\omega_p$ the Hamiltonian reads
\begin{equation} \label{equ: Hamiltonian in 1-2 basis}
\hat{\mathcal H} = \sum_{i=1,2} (-\Delta \hat a_i^\dag \hat a_i + U\hat a_i^\dag \hat a_i^\dag \hat a_i \hat a_i) - J(\hat a_1^\dag \hat a_2 + \hat a_1 \hat a_2^\dag) + F(\hat a_1 - \hat a_2 + (\hat a_1 - \hat a_2)^\dag).
\end{equation}
Here, $\hat a_i$ is the bosonic annihilation operator of the $i$-mode,  $\Delta = \omega_p - \omega_c$ is the frequency detuning between the pump frequency and the resonant frequencies $\omega_c$ of the two modes, $U$ is the interaction strength, $J$ is the coupling between the two modes, $F$ is the pump amplitude, and $\gamma$ is the decay rate.

\begin{figure}[t]
\includegraphics[width=1\linewidth]{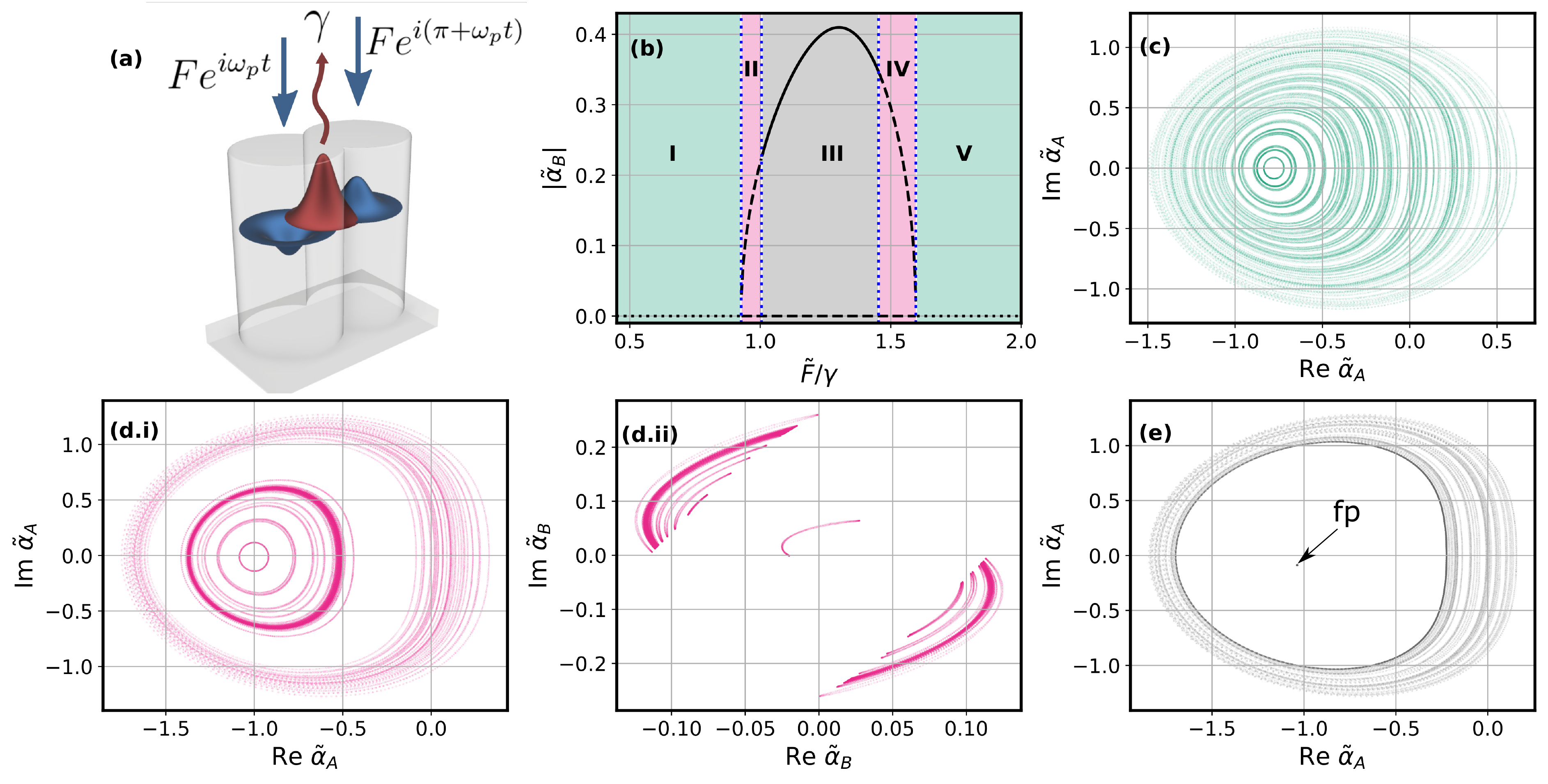}
\caption{{\bf Semiclassical symmetry breaking.} (a) Artistic representation of bonding (red) and antibonding (blue) bosonic wavefunctions inside two coupled microcavity pillars.
(b) Rescaled order parameter $|\tilde \alpha_B|$ as a function of the rescaled pump amplitude. Different regions denoted by capital Roman numerals are separated by blue vertical dashed lines and are discussed in the main text. Black continuous (dotted) lines indicate attractive (non-attractive stable) fixed points, while black dashed lines depict repulsive fixed points. 
(c, d, e) Phase space portrait for pump ampltiudes $\tilde F/\gamma = 0.5$, $0.95$, and $1.2$, in regions I, II and  III, respectively. The parameters read $\Delta/\gamma = 0.8$, $J/\gamma = 1.1$, and $\tilde U/\gamma=1$. We use the same parameter values throughout the manuscript.}
\label{fig: symm break semiclassical}
\end{figure}

To better appreciate the effects the drive and the dissipation have over the two modes, we can rewrite the Lindblad equation in terms of bonding and antibonding modes $\hat a_B = (\hat a_1 + \hat a_2)/\sqrt{2}$ and $\hat a_A = (\hat a_1 - \hat a_2)/\sqrt{2}$, respectively, obtaining $\partial_t \hat \rho = -i[\hat{\mathcal H} , \hat \rho] + 2\gamma \mathcal{D} [\hat a_B](\hat \rho)$ with
\begin{equation} \label{equ: bonding and antibonding Hamiltonian}
\begin{split}
 \hat{\mathcal H} = &(-\Delta-J) \hat a_B^\dag \hat a_B + (-\Delta + J) \hat a_{A}^\dag \hat a_{A} + \sqrt{2}F(\hat a_{A}^\dag + \hat a_{A}) + \frac{U}{2}\left[ \sum\limits_{k=B,A}(\hat a^\dag_k \hat a^\dag_k \hat a_k \hat a_k) \right. \\
&+ \hat a_B^\dag \hat a_B^\dag \hat a_A \hat a_A + \hat a_B \hat a_B \hat a_A^\dag \hat a_A^\dag + \ 4 \hat a_B^\dag \hat a_B \hat a_A^\dag \hat a_A \Bigg].
\end{split}
\end{equation}
In this new basis we can note that only the bonding mode dissipates and only the antibonding mode is driven, while the interaction ($U$) couples both modes. This bosonic dimer (although with local dissipation) has been engineered using exciton-polaritons in microcavity pillars \cite{Galbiati_PRL2012}. We illustrate our system in the context of micropillars in figure \ref{fig: symm break semiclassical}(a) showing an \textit{artistic representation} of how the bonding and antibonding modes look in the two coupled pillars. The bonding (red colored) and antibonding (blue colored) modes resemble the symmetric and antisymmetric  wavefunctions, respectively, of the hydrogen molecule. Semiconductor microcavities are not the only platform available to study dissipative and interacting bosonic modes; circuit QED or optomechanical devices are also suitable. The drive we consider in this work can be achieved in any of these three platforms by using two coherent drives, one for each one of the 1 and 2 bosonic modes, with a $\pi$-phase difference, as illustrated in figure \ref{fig: symm break semiclassical}(a). Our nonlocal dissipation could be engineered in microcavity pillars by deliberately introducing defects in the overlap region of the two pillars in order to increase nonradiative losses in the bonding mode. Alternatively, it could be engineered in circuit QED devices by coupling two resonant modes to a microwave resonator at the same position, such that the amplitude and phase of the linear coupling between the resonator and each mode are the same. Moreover, reservoir-engineered nonlocal dissipation has already been used in optomechanical circuits to achieve non-reciprocity \cite{fang2017dissip_coupling_optomechanics, bernier2017nonreciprocal}.

There is an exact semiclassical limit for the system governed by equation (\ref{equ: Lindblad equation}) in which the occupation numbers of modes 1 and 2 diverge and the system's criticality becomes manifest \cite{Casteels-Ciuti_PRA2017, Casteels-Fazio-Ciuti_PRA2017}. It is defined by taking $F \to +\infty$ while keeping $F\sqrt{U}$ fixed. Is thus convenient to introduce a scaling parameter $N$ to define $F=\tilde F \sqrt{N}$ and $U=\tilde U/N$. From (\ref{equ: Lindblad equation}) we obtain the semiclassical mean-field equations
\begin{subequations} \label{equ: bonding and antibonding SC equations}
\begin{align}
i\partial_t \tilde \alpha_B &= (-\Delta - J - i\gamma)\tilde \alpha_B + \tilde U(|\tilde \alpha_B|^2 \tilde \alpha_B + \tilde \alpha_A^2 \tilde \alpha_B^* + 2|\tilde \alpha_A|^2\tilde \alpha_B) \label{equ: bonding SC} \\
i\partial_t \tilde \alpha_A &= (-\Delta + J)\tilde \alpha_A + \tilde U (|\tilde \alpha_A|^2 \tilde \alpha_A + \tilde \alpha_B^2 \tilde \alpha_A^* + 2|\tilde \alpha_B|^2 \tilde \alpha_A) + \sqrt{2}\tilde F, \label{equ: antibonding SC}
\end{align}
\end{subequations}
where we have defined $\tilde \alpha_{A,B} = \alpha_{A,B}/\sqrt{N}$. We can see that these equations are scale invariant and, in particular, they are exact ($\mean{\hat a_{1,2}} = \alpha_{1,2}$) in the weak-coupling $N\to +\infty$ limit, which we study in this work. We will consider parameter values for which equations (\ref{equ: bonding and antibonding SC equations}) are not amenable to perturbative expansions (see \cite{bose1989Anharmonic} for instance), as all energy scales will be equally important. Note that in the laboratory, the interaction strength $U$ is not easily controlled, yet is usually weaker than all other energy scales. Therefore, a large $N$ limit is achieved solely by increasing the pump amplitude up to the level where the modes population become large enough such that the effective interaction energy $U\langle \hat a_i^\dag \hat a_i \rangle$ is relevant. For this reason $F$ is the only parameter we vary throughout this work.

In a previous work \cite{Lledo2019_BHD} we considered the same model but with a pump acting over the bonding instead of the antibondig mode. There we found that the time-translational symmetry of the steady state was broken and it was accompanied by a first-order dissipative phase transition in the form of bistability. This means there was a region of parameter space where one could see long-lived oscillations with one of two different frequencies, depending on the initial conditions. In the current work the symmetries are different, so bistability is now replaced by a second-order phase transition due to the breaking of a $\mathbb Z_2$ symmetry, as we explain in the following. 

\subsection{Symmetries}

Our system has a discrete (also dubbed weak, see \cite{Buca2012_symmetry}) $\mathbb{Z}_2$ symmetry $\hat a_1 \leftrightarrow -\hat a_2$ (or $\tilde \alpha_1 \leftrightarrow -\tilde \alpha_2$) described by the bonding parity operator $\hat P =(-1)^{\hat a_B^\dag \hat a_B} =\sum\limits_{n_1, n_2=0}^\infty (-1)^{n_1+n_2} \ket{n_1, n_2}\bra{n_2,n_1}$ (written in the Fock basis of the 1,2 modes). Note that due to the coherent drive, there is no $U(1)$ phase invariance.

For any finite value of $N$, the dynamical equation (\ref{equ: Lindblad equation}) has a unique steady state (i.e., time-independent) which is symmetric $\hat \rho_\text{ss} = \hat P \hat \rho_\text{ss} \hat P^\dag$. However, both the $\mathbb{Z}_2$ symmetry and the time-translational symmetry of the steady state can be broken in the $N \to +\infty$ limit where the number of bosons in the system diverge. Naturally, then, our order parameters to witness spatial and time symmetry breakings in the semiclassical limit should be $\lim_{t\to\infty} \tilde \alpha_B(t) = \lim_{t,N\to \infty} N^{-1/2} \Tr[\hat a_B \hat \rho(t)]$ and any time-dependent function $f(\tau) = \lim_{t,N\to \infty} \Tr[\hat O \hat \rho(t+\tau)]$, respectively. If $f(t)$ is periodic, then the system would be in a time-crystalline phase. In the next section we will show that we find periodic oscillations for all values of the pump ampltiude $\tilde F$, but the $\mathbb Z_2$ symmetry is broken only for a particular region.

\section{Mean-field semiclassical dynamics and symmetry breakings}

In this section we analyse the dynamical behaviour of our semiclassical model (\ref{equ: bonding and antibonding SC equations}). We look for fixed points and their stability, as well as the formation of limit cycles. In the first part we present the numerical results we obtain by solving (\ref{equ: bonding and antibonding SC equations}) and then we give an analytical explanation. But first, since the nomenclature for nonlinear dynamics varies from one physics community to another, we first give a brief summary. 

The fixed points are the stationary solutions, i.e., $\partial_t \tilde \alpha_{A,B}^\text{fp}=0$, and their local stability is deduced from the eigenvalues of the Jacobian matrix obtained linearising the equations around them. If we express (\ref{equ: bonding and antibonding SC equations}), and their complex conjugates, in vector notation as $\partial_t \bm{\alpha} = \bm{G}(\bm \alpha)$ where $\bm{\tilde \alpha} = (\tilde \alpha_B, \tilde \alpha_B^*, \tilde \alpha_A, \tilde \alpha_A^*)^T$, the expansion $\bm \alpha \to \bm \alpha^\text{fp} + \bm \delta$ for $\bm \delta$ the fluctuations vector leads us to the linear equation $\partial_t \bm \delta = \bm M \bm \delta$ where $\bm M = (\partial \bm G/\partial \bm \alpha)|_\text{fp}$ is the Jacobian matrix. Depending on the real part of the eigenvalues, the fixed point can be (locally) attractive, stable but non-attractive, or repulsive, corresponding to having all eigenvalues negative, at least one equal to zero, or at least one positive, respectively.
A limit cycle is a periodic orbit in phase space; when a trajectory enters a limit cycle it remains there forever.

\subsection{Numerical results}

We find five regions in parameter space which are shown in figure \ref{fig: symm break semiclassical}(b), highlighted in different colors and annotated using Roman numerals. We plot the rescaled order parameter $|\alpha_B|/\sqrt{N}$ of the $\mathbb Z_2$ symmetry as a function of the rescaled driving amplitude $F/\sqrt{N}\gamma$, showing that in regions II, III, and IV the $\mathbb Z_2$ symmetry is broken. The time-translational symmetry is broken in all regions. In the following we explain the dynamics in each one of them.

In regions I and V there is a single non-attractive fixed point (dotted black line in figure \ref{fig: symm break semiclassical}(b)) which preserves the symmetry. This fixed point is never reached; all the trajectories go into a family of limit cycles on the antibonding mode alone, as the bonding mode amplitude goes to zero. These can be seen in panel (c) of the same figure, where we show a \textit{phase space portrait} for $\tilde F=0.5$ in region I. For panel (b) [as well as for (c), (d), and(e)] we allow 100 random initial conditions to evolve until they have become stationary and then we sample $\tilde \alpha_{A,B}(t)$ with a rate $(\gamma \Delta t)^{-1} = 10^{-2}$.

In regions II and IV there are three repulsive fixed points (dashed black line); one of them preserves the symmetry and two of them belong to a symmetry-related pair ($\tilde \alpha_B \leftrightarrow -\tilde \alpha_B$) breaking the symmetry. Here there is also a family of symmetry-preserving limit cycles revolving around the fixed point with $\alpha_B = 0$. Additionally, there are symmetry-breaking attractive limit cycles: in the low end of region II, a pair of them emerge from the fixed point as the pump amplitude rises, converging eventually to the fixed points with symmetry-breaking when region III is reached (see below). The same happens in regions IV in reverse. The different limit cycles can be seen in figures \ref{fig: symm break semiclassical}(d.i) and (d.ii), where we plot a phase space portrait for $\tilde F=0.95$. Note we now plot the phase space of both modes as the bonding mode does not vanish.

Finally, in region III, the two symmetry-breaking repulsive fixed points of II and IV become attractive. There are only symmetry-preserving limit cycles in this region and they revolve around the repulsive fixed point with $\alpha_B=0$. The fixed point (indicated by an arrow) and the limit cycles can be seen in panel (e), where we plot a phase portrait for $\tilde F=1.2$. We do not show the bonding mode in this case, as it goes to zero (for the limit cycles) or to the finite fixed point value indicated by the order parameter in panel (b).

We can connect the image in figure \ref{fig: symm break semiclassical}(a) with the dynamics just described to have a pictorial understanding. Succinctly we can say: In regions I and V the red (symmetric) mode is not occupied and the blue (antisymmetric) mode always oscillates; in regions II and IV both red and blue modes always oscillate; and in region III either the blue mode oscillates and the red mode is empty or both modes are occupied without any oscillations.

\begin{figure}[t]
\includegraphics[width=1.0\linewidth]{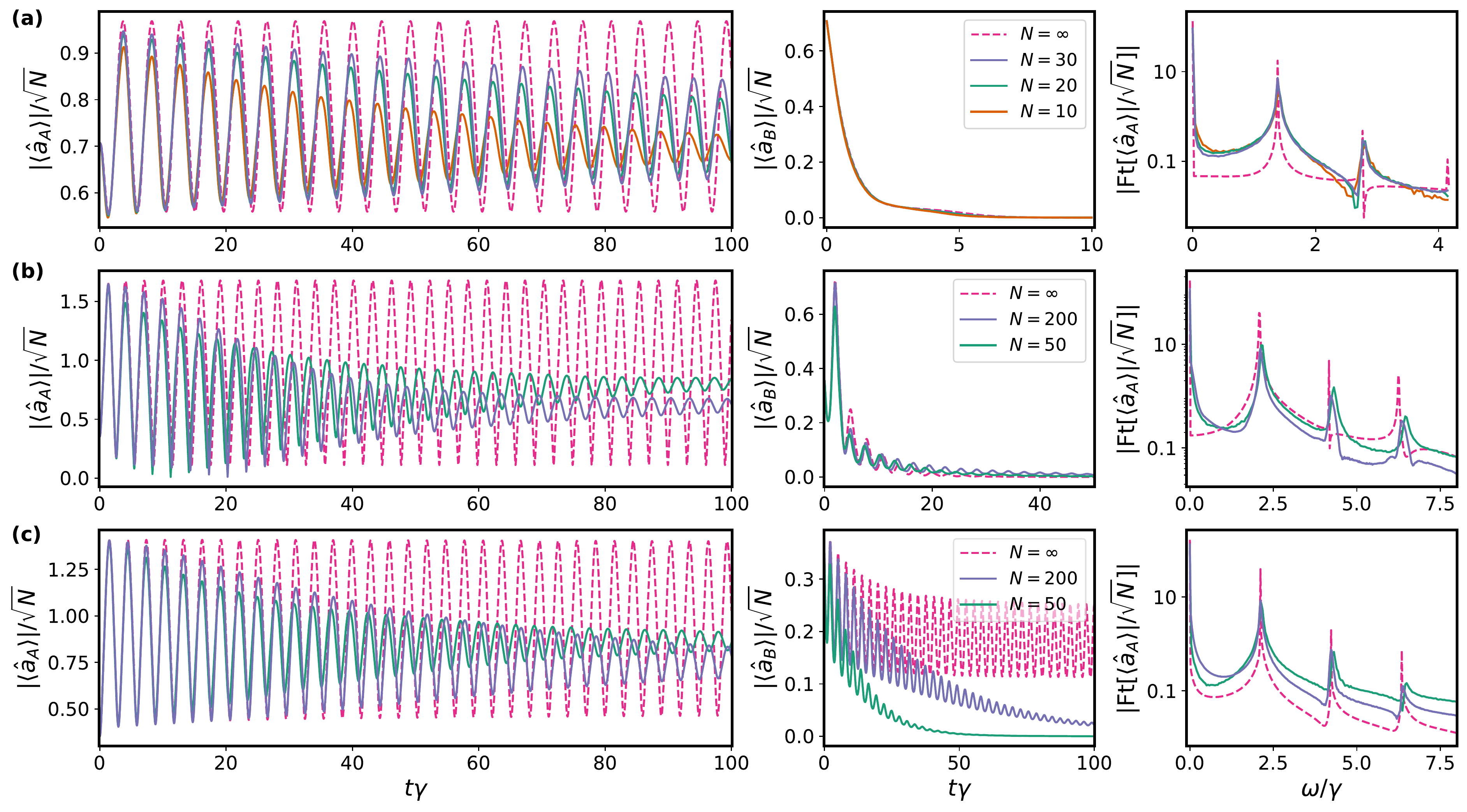}
\caption{{\bf Emergence of limit cycles.} Amplitude of the bonding and antibonding modes as a function time and the Fourier spectrum of the antibonding mode for different values of $N$. (a) Symmetry preserving oscillations for low pump $F/(\sqrt{N}\gamma) = 0.5$, in region I. (b, c) Symmetry-preserving and symmetry-breaking oscillations, respectively, for two different initial conditions and $F/(\sqrt{N}\gamma) = 0.95$ in region II. The initial conditions are coherent states with $\alpha_2=0$ and $\alpha_1/\sqrt{N} = -1$, $0.5$, and $-0.5$ for (a), (b), and (c), respectively. The results for $N=10,20,30$, and $50$ were obtained averaging over $5\times10^3$ quantum jump trajectories, and for $N=200$, averaging over $10^4$ TWA trajectories. 
For the Fourier spectrum, the antibonding mode was sampled with $\Delta t\gamma=10^{-2}$ over total time $T=300$ and $150$ for the simulations with $N\geq 200$ and $N<200$, respectively. The rest of the parameters are $(\Delta, J, U/N)= (0.8, 1.1, 1)\gamma$, same as in figure \ref{fig: symm break semiclassical}.}
\label{fig: limit cycles}
\end{figure} 

In figure \ref{fig: limit cycles} we show examples of limit cycles for regions I and II by plotting the modes' amplitudes as a function of time. For the moment we just focus on the curves for $N=\infty$, which correspond to the semiclassical case. In panel (a), for low pump ($\tilde F/\gamma = 0.5$), we can see that the limit cycle is reached when the bonding mode goes to zero on a short time scale. In panels (b) and (c) we show the same quantities for a pump amplitude $F/\sqrt{N}\gamma = 0.95$, inside region II. In panel (b) the initial condition is such that a symmetry-preserving limit cycle is reached, while in (c), it is such that a symmetry-breaking limit cycle is obtained. On the right hand side of the figure, we show the Fourier transforms of the antibonding mode $\alpha_A/\sqrt{N}$ and we observe there are equidistant frequency peaks akin to frequency combs. This implies the corresponding periods are commensurate with each other, meaning a single common period governs the oscillations. Nevertheless, the frequency difference between consecutive peaks has a nontrivial dependence on all parameters. Although not shown, we note that all the frequency peaks can also be seen in the Fourier spectrum of the dynamics after the transient, when the system is in a fully periodic motion. Moreover, the same frequencies are obtained from the amplitude or phase oscillations.

\subsection{Analysis}

Some intuition about the dynamical behaviour can be gained by noting that, if $J-\Delta > 0$, which is our case, a fixed point of (\ref{equ: bonding and antibonding SC equations}) is given by
\begin{equation} \label{equ: fixedpoint}
\begin{split}
\tilde \alpha_B^\text{fp} &= 0, \\
\tilde \alpha_A^\text{fp} &= \left( -\frac{\tilde F}{\sqrt{2} \tilde U} + \sqrt{\frac{\tilde F^2}{2 \tilde U^2} + \left(\frac{J-\Delta}{3\tilde U}\right)^3} \right)^{1/3} - \left(\frac{\tilde F}{\sqrt{2} \tilde U} + \sqrt{\frac{\tilde F^2}{2 \tilde U^2} + \left(\frac{J-\Delta}{3\tilde U}\right)^3} \right)^{1/3}
\end{split}
\end{equation} 
where $\tilde \alpha_A^\text{fp}$ is real and negative and the solution of the depressed cubic equation $\tilde U (\tilde \alpha_A^\text{fp})^3 + (J-\Delta) \tilde \alpha_A^\text{fp} + \sqrt{2} \tilde F = 0$. This is the symmetry-preserving fixed point found for all $\tilde F$ in the numerical analysis and shown in figure \ref{fig: symm break semiclassical}(b). The eigenvalues of the Jacobian matrix which determines its linear stability are given by
\begin{eqnarray}
\lambda^B_{\pm} &=& - \gamma \pm \sqrt{\tilde U^2 (\tilde \alpha_A^\text{fp})^4 - (2 \tilde U (\tilde \alpha_A^\text{fp})^2 - \Delta -J)^2}, \\
\lambda^A_{\pm} &=& \pm i \sqrt{(J-\Delta + 2 \tilde U (\tilde \alpha_A^\text{fp})^2)^2- \tilde U^2 (\tilde \alpha_A^\text{fp})^4}. \label{equ: SC frequency}
\end{eqnarray}
The eigenvalues $\lambda_A^{\pm}$ are purely imaginary meaning (\ref{equ: fixedpoint}) cannot be attractive. Instead, a family of limit cycles can be reached: whenever $\tilde \alpha_B=0$ the two modes decouple, and the dynamical equation of $\tilde \alpha_A(t)$ corresponds to the semiclassical equation of a coherently driven, dissipationless nonlinear harmonic oscillator
\begin{equation} \label{equ: antibonding equ after decoupling}
i\partial_t \tilde \alpha_A \approx (-\Delta + J +\tilde U |\tilde \alpha_A|^2) \tilde \alpha_A + \sqrt{2} \tilde F.
\end{equation}
The limit cycle attained depends on all parameters and the value of $\tilde \alpha_A(t)$ at the time when $\tilde \alpha_B$ vanishes. A similar effective decoupling mechanism has been studied in \cite{Lledo2019_BHD}.

There is an instability signalled by $\text{Re}[\lambda^B_+] > 0$ where two symmetry-breaking fixed points emerge. This inequality is satisfied in a region where the following two conditions are fulfilled:
\begin{equation} \label{equ: instability condition}
J + \Delta > \tilde U (\tilde \alpha_A^\text{fp})^2 \qquad \text{and} \qquad \sqrt{\tilde U^2 (\tilde \alpha_A^\text{fp})^4 - (2 \tilde U (\tilde \alpha_A^\text{fp})^2 - \Delta -J)^2} > \gamma.
\end{equation}
For the parameters we use in figure \ref{fig: symm break semiclassical} (same throughout the manuscript) these two inequalities are simultaneously satisfied in the region $\tilde F/\gamma \in [0.927,\,1.596]$, which corresponds to regions II, III and IV.

Interestingly, the bonding mode amplitude need not be zero for the effective decoupling of the two modes. When the amplitude is small ($|\tilde \alpha_B|\ll 1$), equation \ref{equ: bonding and antibonding SC equations}(b) decouples from \ref{equ: bonding and antibonding SC equations}(a), resulting in equation (\ref{equ: antibonding equ after decoupling}) which drives the antibonding mode into periodic oscillations. At the same time, equation (\ref{equ: bonding and antibonding SC equations}a) becomes linear in the bonding mode while the antibonding mode acts as a drive giving
\begin{equation}
i\partial_t \tilde \alpha_B \approx (-\Delta - J - i\gamma) \tilde \alpha_B + \tilde U(\tilde \alpha_A(t)^2 \tilde \alpha_B^* + 2 |\tilde \alpha_A(t)|^2 \tilde \alpha_B).
\end{equation}
Note that only the time dependence of $\tilde \alpha_A$ is written explicitly. This is to highlight that, since $\alpha_A(t)$ is periodic, it will cause the bonding mode to oscillate with the same period ---this is, it is acting as a Floquet driving. This explains why in regions II and IV a family of symmetry-breaking limit cycles organise around the unstable fixed points emerging from the instability outlined in (\ref{equ: instability condition}).

We remark on the importance of a pure nonlocal dissipation. If local dissipators in the form of $\kappa \sum_{i=1}^2\mathcal D[\hat a_i](\hat \rho)$ (with small $\kappa\ll \gamma$) were to be added to (\ref{equ: Lindblad equation}), the oscillations in the bonding and antibonding modes would be damped for all values $\tilde F$. Even though the system would preserve the $\mathbb Z_2$ symmetry, one would be able to observe long-lived oscillations only up to a time $\sim \kappa^{-1}$.

Having given an analytical explanation for the phase space of our system in the $N\to +\infty$ limit, we remark that an analysis that does not go beyond mean-field is incomplete: quantum fluctuations may well destroy the semiclassical limit cycles. A transparent example can be found in \cite{Navarrete2017_PRL}, where a linearisation method is developed to show that quantum fluctuations smear out the semiclassical limit cycle of the Van der Pol oscillator. In order to prove the robustness against quantum fluctuations, we proceed in the next section by solving our system exactly deep in the quantum regime ($N=1$) and carrying out a 'finite-size' expansion in terms of the scaling parameter $N$.


\section{Quantum dynamics}

For our numerical calculations in this section, we have appropriately truncated the Fock space in the 1-2 basis ensuring convergence in the results for each value of the scaling parameter $N$. For the time evolution of expectation values, we solve (\ref{equ: Lindblad equation}) using a photon-counting unravelling \cite{Carmichael_lecture_notes} of the master equation and averaging over $5\times10^3$ quantum jump trajectories, recovering to an excellent accuracy the full dynamics of the density matrix. We also use the Wigner phase space representation and the Truncated Wigner Approximation (TWA) \cite{Carmichael_lecture_notes} ---we invite the reader to see \ref{appendix: TWA} for details.

By studying the time evolution of the rescaled expectation values $\mean{\hat a_{A,B}}(t)/\sqrt{N}$ for increasing values of $N$, we gain insight on both symmetry breakings we are interested in. Firstly, the emergence of periodic oscillations in $|\mean{\hat a_A}|(t)/\sqrt{N}$ for all the pump amplitudes we consider would highlight the time-translation symmetry breaking. Secondly, if the $\mathbb{Z}_2$ symmetry is to be broken for an $\tilde F$ window (i.e., regions II-IV), the response of $|\mean{\hat a_B}|(t)/\sqrt{N}$ should provide evidence for critical slowing down in this region due to the necessary closing of the Liouvillian eigenvalue gap \cite{CiutiSpectralTheo} (more to this below). 

We compare the previously shown semiclassical limit cycles with the dynamics of the quantum regime in figure \ref{fig: limit cycles}. It clearly depicts the emergence of periodic oscillations in the antibonding mode as $N$ is increased. Recall that panel (a) corresponds to a low pump amplitude in region I, while panels (b) and (c) correspond to region II with different initial conditions. Looking at the oscillations in the bonding mode in (b) and (c), we can see that also in the quantum regime a symmetry-preserving or symmetry-breaking limit cycle is approached with increasing $N$. Furthermore, we can observe in the Fourier spectrum of figures \ref{fig: limit cycles} (a), (b), and (c) that the long-lived oscillations in the quantum regime have frequency peaks matching those of the semiclassical limit. In particular, this proves that the period of the oscillations is robust against quantum fluctuations. For $N$ up to 50 we have used quantum jump trajectories while for the very high $N=200$ we have used the TWA which is a very good approximation for weak interactions ($U=\tilde U/N = 0.005$). 

Thanks to the linearity of the Lindblad equation we can expand its formal solution in eigenmodes of the superoperator $\mathcal L$ as
\begin{equation} \label{equ: expansion of lindblad equ}
\hat \rho(t) = e^{\mathcal L t}(\hat \rho(0)) = \sum_{j\geq 0} c_j e^{\lambda_j t} \hat \rho_j = \hat \rho_\text{ss} + \sum_{j\geq 1} c_j e^{\lambda_j t} \hat \rho_j,
\end{equation}
where $\hat \rho_j$ are the eigenmodes, $\lambda_j$ the complex eigenvalues, and $c_j$ coefficients that depend on the initial condition. As our notation suggests, $\lambda_0 = 0$ is associated with the steady state eigenmode $\hat \rho_0 = \hat \rho_\text{ss}$. The other eigenvalues have negative real part (verified numerically), hence $|\text{Re }\lambda_j|$ correspond to decay rates of the transient modes, while $\text{Im }\lambda_j$ are frequencies. Since the steady state preserves the symmetry of $\mathcal L$, this picture suggests that to have a phase transition at least one non-zero eigenvalue must vanish in the $N\to + \infty$ limit (or in a thermodynamic limit, in general). This is analogous to the gap closing in a second-order phase transition of a closed quantum system \cite{Sachdev}.

In order to obtain the eigenvalues of a general Liouvillian superoperator $\mathcal L$, one normally proceeds by first writing it as a matrix, $\check{\mathcal L}$, and then diagonalizing it. In our case, thanks to the discrete $\mathbb Z_2$ symmetry, the $\check{\mathcal{L}}$ obtained from (\ref{equ: Lindblad equation}) decomposes into two block matrices as
\begin{equation} \label{equ: subspaces separation}
\check{\mathcal L} = \check{\mathcal L}^+ \oplus \check{\mathcal L}^- = \begin{bmatrix}
\check{\mathcal L}^+ & \mathbf 0 \\
\mathbf 0 & \check{\mathcal L}^-
\end{bmatrix},
\end{equation}
where $(\pm)$ refers to the two eigenspaces corresponding to eigenvalues $\pm 1$ of the symmetry operator $\check{\mathcal P} \equiv \hat P \otimes \hat P^*$ that commutes with $\check{\mathcal L}$. Details can be found in \ref{apppendix: Symmetry}. The usefulness of this is twofold. Firstly, it reduces the computational cost of finding the eigenvalues. Secondly, eigenmodes in the $(+)$-subspace are $\mathbb Z_2$-symmetric while eigenmodes in the $(-)$-subspace are antisymmetric. This means we should be able to find evidence of the time-translational symmetry breaking in the spectrum of $\check{\mathcal L}^+$ alone, and of the $\mathbb{Z}_2$ symmetry breaking in the spectrum of $\check{\mathcal L}^-$.

\begin{figure}[t]
\includegraphics[width=1.0\linewidth]{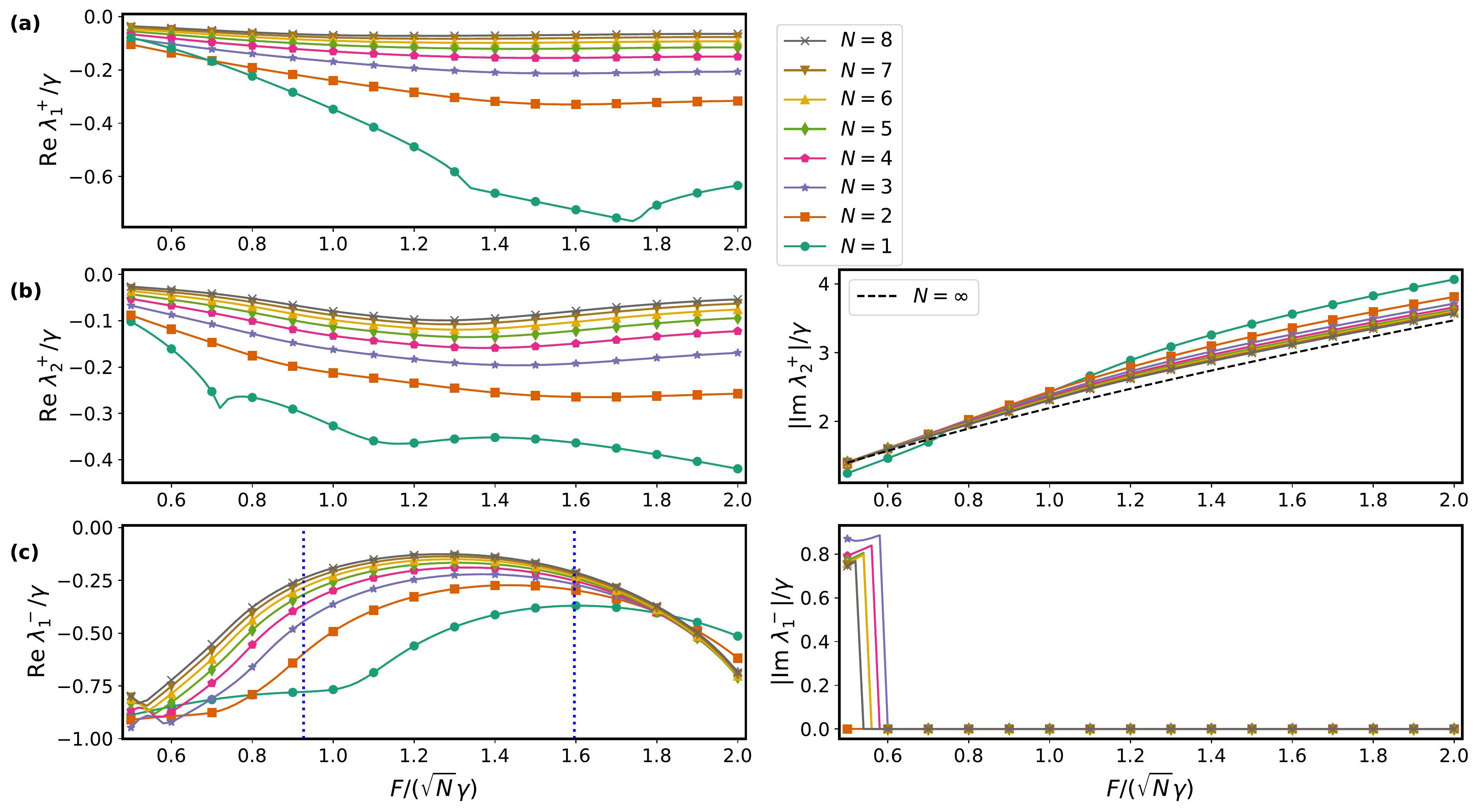}
\caption{{\bf Liouvillian's eigenvalues} with smallest absolute real part as a function of $F/(\sqrt{N}\gamma)$ for different values of $N$. (a),(b) The two smaller eigenvalues of $\check{\mathcal L}^+$. $\lambda_1^+$ is real and $\lambda_2^+$ is complex, with its imaginary part approaching the semiclassical frequency (\ref{equ: SC frequency}) (black dashed line) as $N$ is increased. (c) The smaller eigenvalue of $\check{\mathcal L}^-$. Its real part approaches zero by increasing $N$ in a region where its imaginary part vanishes. The region between vertical dashed blue lines corresponds to regions II, III, and IV where the semiclassical analysis predicts breaking of the $\mathbb Z_2$ symmetry. The markers are just a guide; each curve contains 76 points.}
\label{fig: Liouvillian spectrum}
\end{figure} 

In figure \ref{fig: Liouvillian spectrum} we show the non-zero eigenvalues of $\mathcal L$ with smallest absolute real part as a function of the pump amplitude for different values of $N$. In panel (a), we show an eigenvalue ($\lambda_1^+$) of $\check{\mathcal L}^+$ which is purely real and goes to zero as $N$ increases. In panel (b), for the same symmetry subspace, we show a second eigenvalue ($\lambda_2^+)$ which is complex. When $N$ is increased, its real part tends to zero while its imaginary part converges to a finite value. This eigenvalue is responsible for the emergent oscillations for all the pump values we consider. In order to appreciate this, we have included in panel (b) the frequency predicted from the semiclassical linearised equations, i.e., from (\ref{equ: fixedpoint}) and (\ref{equ: SC frequency}). Clearly $|\text{Im } \lambda_2^+|$ approaches the semiclassical frequencies.

The smallest eigenvalue ($\lambda_1^-$) of $\check{\mathcal L}^-$ tends to zero with increasing $N$ in the region where the semiclassical analysis predicts the broken $\mathbb Z_2$ phase. This is shown in \ref{fig: Liouvillian spectrum}(c), where we have delimited between vertical blue dashed lines regions II, III and IV (in the plot for the real part). The real part of the eigenvalue has an inverted parabolic shape approaching zero with increasing $N$, while the imaginary part is zero throughout this region. The asymptotic vanishing of this eigenvalue explains the critical slowing down in the convergence to the steady-steady expectation values of non-symmetric operators, as shown for the bonding mode in figure \ref{fig: limit cycles}(c). Figure \ref{fig: limit cycles}(b) does not show the same critical slowing down in the bonding mode in spite of sharing the same pump amplitude. This can be understood by recalling that the coefficients $c_j$ in (\ref{equ: expansion of lindblad equ}) depend on the initial conditions; in the latter case the initial state does not couple to the eigenmode associated to $\lambda_1^-$.

\begin{figure}[t]
\includegraphics[width=1.0\linewidth]{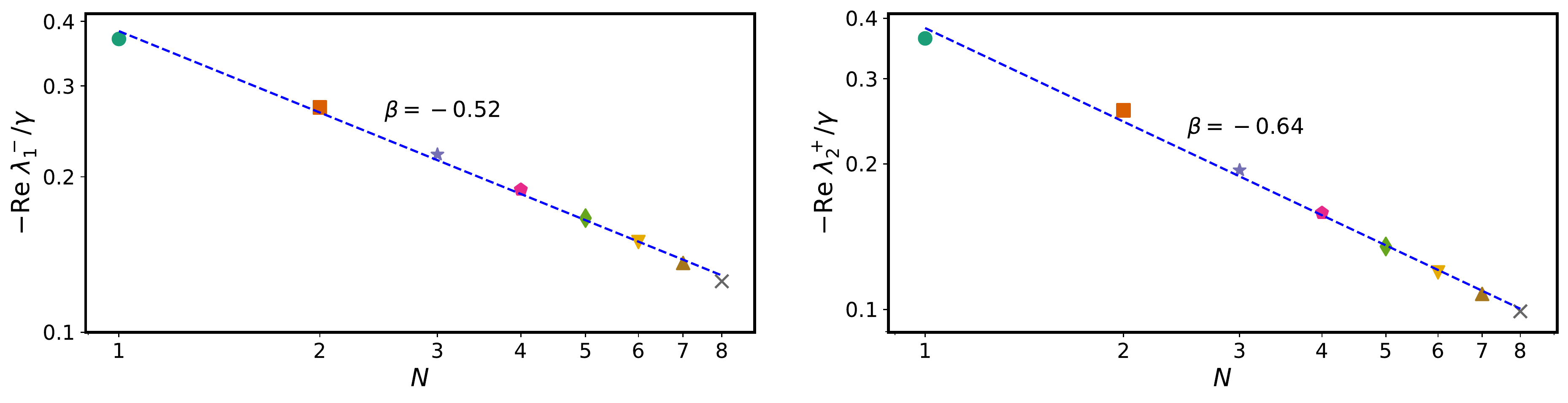}
\caption{{\bf Algebraic scaling of the eigenvalues.} Re $\lambda_1^-$ and Re $\lambda_2^+$ as a function of $N$ in log-log scale. For each $N$, we take the largest value of the curves for Re $\lambda_1^-$ in the left panel of figure \ref{fig: Liouvillian spectrum}(c). At the same values of pump amplitude, we extract the Re $\lambda_2^+$. In the left (right) panel we see the eigenvalue show a dependence $\lambda \propto N^\beta$ with $\beta = -0.52$ ($\beta = -0.64$) (fit, dashed blue).}
\label{fig: critical scaling}
\end{figure} 

Figure \ref{fig: critical scaling} shows the scaling of $\text{Re }\lambda_1^-$ and $\text{Re }\lambda_2^+$ with $N$ in a log-log scale. For Re $\lambda_1^-$, we take the largest value of each curve shown in figure \ref{fig: Liouvillian spectrum}(c), and we extract Re $\lambda_2^+$ for the same values of the pump amplitude. We see that both eigenvalues are well fitted by an algebraic scaling $\lambda \propto N^\beta$ with $\beta = -0.52$ for $\lambda_1^-$ and $\beta = -0.64$ for $\lambda_2^+$, showing that their real part vanish in the $N \to +\infty$ limit.

\begin{figure}[t]
\includegraphics[width=1.0\linewidth]{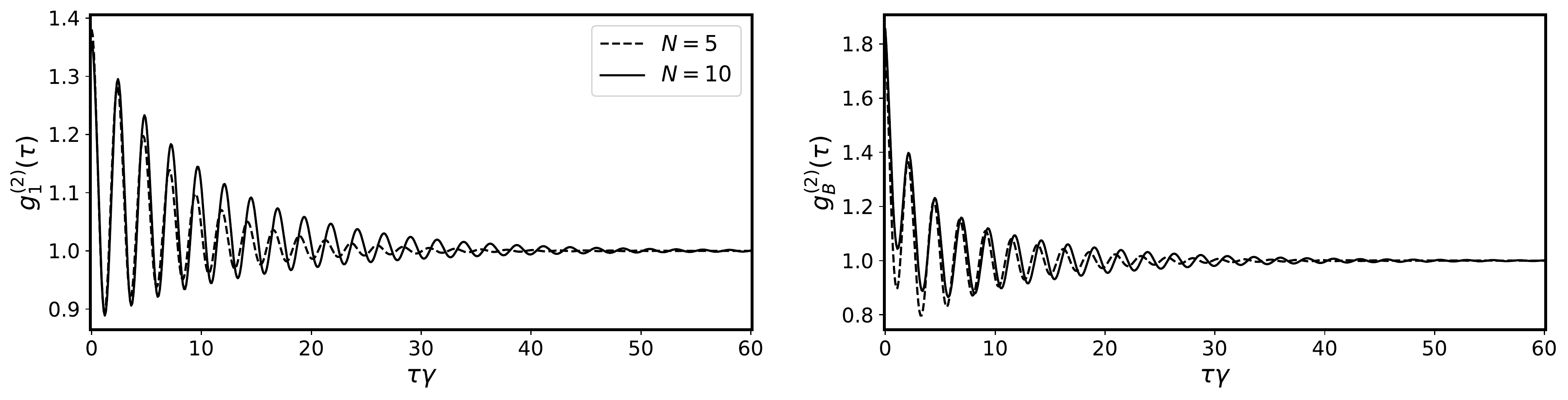}
\caption{{\bf Second-order coherence} of bosonic modes 1 (left) and $B$ (right) in region III ($F/(\sqrt{N}\gamma) = 1.2$) and for two values of $N$. The first detection is done in the steady state.}
\label{fig: g2-time-delayed}
\end{figure}

In region III the semiclassical prediction is not completely accurate. It predicts there is either a steady state with $\mathbb Z_2$ broken or a limit cycle only in the antibonding mode. Yet from figures \ref{fig: Liouvillian spectrum} and \ref{fig: critical scaling} one can deduce limit cycles will always be found in both modes throughout region III. Even though one could fine tune the initial condition such that a symmetric or non-symmetric steady state is reached ---in the $N\to\infty$ limit when the gaps are closed--- any perturbation or time-delayed two-point measurement on the system would be enough to make the bonding and antibonding occupations oscillate forever. To illustrate this point, we show in figure \ref{fig: g2-time-delayed} the second-order coherence of the bosonic modes $\hat a_1$ (left panel) and $\hat a_B$ (right panel) for $\tilde F=1.2$ in region III, and two values of $N$. In the long time limit, the coherence is given by 
\begin{equation}
g^{(2)}_{j}(\tau) = \lim\limits_{t\to\infty} \frac{\langle \hat a_j^\dag(t) \hat a_j^\dag(\tau+t) \hat a_j(\tau+t) \hat a_j(t) \rangle}{\langle \hat a^\dag_j(t) \hat a_j(t) \rangle^2} = \frac{\Tr[\hat a_j^\dag \hat a_j e^{\tau \mathcal L}(\hat \rho'_j)]}{\Tr[\hat a^\dag_j \hat a_j \hat \rho_\text{ss}]}, 
\end{equation}
where $j$ can be $1,2,A$, or $B$, and we have defined $\hat \rho'_j = \hat a_j \hat \rho_\text{ss} \hat a_j^\dag/\Tr[\hat a^\dag_j \hat a_j\hat \rho_\text{ss}]$. The last equality in the above equation emphasises the following: the first detection transforms $\hat \rho_\text{ss}$ into $\hat \rho'_j$, which is then evolved up to a time $\tau$ where the second detection occurs. Noting that $\hat \rho'_j$ can be expanded as a linear combination of $\mathcal L$'s eigenmodes, as in (\ref{equ: expansion of lindblad equ}), we can conclude that the long-lived eigenmodes are probed by the measurement \cite{Fink2018}. 

\section{Discussion and outlook}

The time crystal discussed in this work pertains to the class of emerging semiclassical time crystals also discussed in \cite{Keeling2018, Poletti2018, Owen2018, Pal2018, Rey2018, Ueda2018, Lesanovsky2019, Lledo2019_BHD, Kessler2019-PRA, Savona2019, Esslinger-Science2019, chiacchio2019dissipation, Buca-dissipation-2019PRL}. The mechanism behind it can be traced to the effective dynamical decoupling between the bonding and antibonding modes. The key ingredients are that these modes are nonlinearly coupled and that only one of them is explicitly damped (c.f. (\ref{equ: bonding and antibonding Hamiltonian})). The other (non-damped) mode can evolve autonomously when the population in the damped one is small. We note that this behaviour is closely related to the one found in frequency combs in the \textit{weak-lasing} regime of dissipatively coupled condensates \cite{RuboPRL15}. The mean-field model for the two dissipatively-coupled Bose-Einstein condensates considered in that work, can also be obtained by the semiclassical limit of the Lindblad master equation
\begin{equation}
\begin{split}
\partial_t \hat \rho = &-i\left[ \sum_{i=1}^2 \left( \omega_i \hat a_i^\dag \hat a_i + \frac{\alpha}{4} \hat a_i^\dag \hat a_i^\dag \hat a_i \hat a_i \right) -\frac{J}{2}(\hat a_1 \hat a_2^\dag + \hat a_1^\dag \hat a_2), \hat \rho \right] \\
&+\gamma \mathcal D[\hat a_1 + \hat a_2] (\hat \rho) + \sum_{i=1}^2 (\Gamma - \gamma)\mathcal D[\hat a_i] (\hat \rho) + W \mathcal D[\hat a_i^\dag] (\hat \rho),
\end{split}
\end{equation}
which considers incoherent ($W$) instead of coherent pump. The dissipative coupling is crucial, and the limit cycles are found when the population in the two modes are small and the nonlinearity $\alpha$ becomes inefficient, like in our case.

This phenomenon can be generalised to spatially extended configurations of bosonic modes. In a ring, for instance, one would need that the different linear modes (with well defined angular momentum) have different dissipation rates. Then, coherently pumping one of the linear modes would result in periodic long-lived oscillations in the mode(s) with smallest decay rate(s). If one of them has no decay channel ---like in the case discussed in this work--- then a time crystal would be found.

It would be interesting to relate the limit cycles we find to the mechanism outlined in \cite{Buca2019NatComm}, where an operator $\hat A$ which commutes with the Lindblad operators, $[\hat L_i, \hat A] = [\hat L_i^\dag, \hat A]=0$, is at the same time an eigenoperator of the Hamiltonian $[\hat H, \hat A] = \omega \hat A$. These algebraic conditions are sufficient to show the existence of limit cycles, even in the quantum regime. In our current case, however, one would expect that the frequency $\omega$ (in the $N\to \infty$ limit) is given by the semiclassical frequency in (\ref{equ: SC frequency}), which depends on all parameters including the decay rate $\gamma$. The exact condition for coherent dynamics put forward by Bu\v{c}a et al. in \cite{Buca2019NatComm} gives an $\omega$ depending on Hamiltonian parameters alone. We hypothesise there could be a complementary, and less precise, case to that algebraic condition where $\hat A$, and thus the commutators as well, depend themselves on expectation values, such that only in a thermodynamic limit some of them vanish and a clear indication of coherent dynamics is obtained.


\section{Acknowledgments}
We are thankful to the developers of the Quantum Toolbox in Python (QuTiP) \cite{Qutip1, Qutip2}, as we have used it for most of our calculations in the quantum regime. We thank C. Mc Keever for reviewing the manuscript and Th. K. Mavrogordatos for helpful discussions and early-stage collaboration on this work. C.L. gratefully acknowledges the financial support of the National Agency for Research and Development (ANID)/Scholarship Program/DOCTORADO BECAS CHILE/2017 - 72180352. M.H.S. gratefully acknowledges financial support from QuantERA InterPol and EPSRC (Grant No. EP/R04399X/1 and No. EP/K003623/2). 

\appendix

\section{Truncated Wigner Approximation}\label{appendix: TWA}
In the Wigner representation one can obtain a generalized Fokker-Planck equation for the Wigner quasi-probability distribution
\begin{equation}
W(\alpha_1, \alpha_1^*, \alpha_2, \alpha_2^*, t) \equiv = \frac{1}{\pi^4}\int d^2 z_1 d^2 z_2 \Tr[\hat D(iz^*_1)\otimes \hat D(i z^*_2) \hat \rho(t)]e^{-i(z_1^* \alpha_1^*+z_1 \alpha_1)} e^{-i(z_2^* \alpha_2^*+z_2 \alpha_2)}
\end{equation}
where $\hat D(\alpha) = e^{\alpha \hat a^\dag - \alpha^* \hat a}$ is the displacement operator.

For the system we consider in this work, the \textit{Truncated Wigner Approximation} amounts to neglecting the third-order derivatives
\begin{equation}
\frac{iU}{4}\left(\frac{\partial^2}{\partial \alpha_j^2} \frac{\partial}{\partial \alpha_j^*} - \frac{\partial}{\partial \alpha_j} \frac{\partial^2}{\partial (\alpha_j^*)^2}  \right) W \qquad \text{for $j=1,2$},
\end{equation}
which scale as $\propto N^{-3/2}$, while the first- and second-order derivative terms in the Fokker-Planck equation are made of terms of orders $N$ and $1$.

The truncated Fokker-Planck equation for $W$ can then be mapped into stochastic Langevin equations for $\alpha_i$, yielding
\begin{equation}
\begin{split}
\partial_t \alpha_1 &= [i\Delta - \gamma/2 - 2iU(|\alpha_1|^2-1)]\alpha_1 + (iJ-\gamma/2)\alpha_2 - iF + \sqrt{\gamma/2} \, \eta_1 \\
\partial_t \alpha_2 &= [i\Delta - \gamma/2 - 2iU(|\alpha_2|^2-1)]\alpha_2 + (iJ-\gamma/2)\alpha_1 + iF + \sqrt{\gamma/2} \, \eta_2.
\end{split}
\end{equation} 
$\eta_i$ are complex gaussian noises satisfying $\mean{\eta_i(t)}=0$ and $\mean{\eta_i(t) \eta_j^*(t')} = \delta_{ij} \delta(t-t')$, where the average is over stochastic realizations.

\section{Symmetry of the Lindbladian}\label{apppendix: Symmetry}
We briefly introduce the mapping of the super operator $\mathcal L$ into a matrix, which allows us to build upon common linear algebra knowledge: if two matrices commute, it is possible to separate the matrix into blocks.

In the \textit{doubled Hilbert space} (or \textit{Liouvillian space}) superoperators and operators are mapped into operators and vectors, respectively. Choosing a row-wise reshaping, a superoperator $\mathcal S(\hat \rho) \equiv \hat C_1 \hat \rho \hat C_2$ becomes $\check{\mathcal S} \dket{\rho} = (\hat C_1 \otimes \hat C_2^\top) \dket{\rho}$ where $\top$ is the transpose and $\hat \rho = \sum_{ij} \rho_{ij} \ket{i}\bra{j} \leftrightarrow \dket{\rho} = \sum_{ij} \rho_{ij} \ket{i} \otimes \ket{j}$. 

Our system has the discrete $\mathbb Z_2$ symmetry which can be expressed as $\mathcal L (\hat P \hat \rho \hat P^\dag) = \hat P \mathcal L(\hat \rho) \hat P^\dag$ for any $\hat \rho$, or equivalently, $(\hat P \otimes \hat P^*) \cL = \cL (\hat P \otimes \hat P^*)$. This is, they commute. Since $\hat P$ has eigenvalues $\pm 1$, we can split the Liouvillian into two blocks, and obtain (\ref{equ: subspaces separation}):
\begin{equation}
\cL= \begin{bmatrix}
\cL^+ & \mathbf{0} \\
\mathbf{0} & \cL^-
\end{bmatrix} \equiv
\begin{bmatrix}
\cL_{++,++} & \cL_{++,--} & \mathbf 0 & \mathbf 0 \\
\cL_{--,++} & \cL_{--,--} & \mathbf 0 & \mathbf 0 \\
\mathbf 0 & \mathbf 0 & \cL_{+-,+-} & \cL_{+-,-+} \\
\mathbf 0 & \mathbf 0 & \cL_{-+,+-} & \cL_{-+,-+}
\end{bmatrix},
\end{equation}
where $\cL_{++,++}$ corresponds to projecting onto the eigenspaces of $\hat P$ and $\hat P^*$ with eigenvalue $+1$ on the right and left hand sides of $\cL$, and so on.

\newpage

\bibliography{BHD_NJP_bibliography}

\end{document}